\documentclass[figures,cite,preprint]{epl}

\title{Switching the current through molecular wires \\with Gaussian laser
  pulses} 
\shorttitle{Switching the current through molecular wires} 
\author{Ulrich Kleinekath\"ofer \and GuangQi Li 
  \and Sven Welack  \and Michael Schreiber 
 }
\shortauthor{Ulrich Kleinekath\"ofer \etal}
 \institute{
 Institut f\"ur Physik, Technische Universit\"at Chemnitz, 09107
  Chemnitz, Germany } 

\begin{document}
\maketitle

\begin{abstract}
  The influence of Gaussian laser pulses on the transport through molecular
  wires is investigated within a tight-binding model for spinless electrons
  including correlation. Motivated by the phenomenon of coherent
  destruction of tunneling for monochromatic laser fields, situations are
  studied in which the maximum amplitude of the electric field fulfills the
  conditions for the destructive quantum effect. It is shown that, as for
  monochromatic laser pulses, the average current through the wire can be
  suppressed.  For parameters of the model, which do not show a net current
  without any optical field, a Gaussian laser pulse can establish a
  temporary current. In addition, the effect of electron correlation on the
  current is investigated.
\end{abstract}

\section{Introduction}
In recent years many groups have been working on making the vision of
molecular electronics reality \cite{nitz03a,ghos04}. This bottom-up
approach for electronic devices has certain advantages over standard
top-down approaches which are mainly used these days. In molecular
electronics the transport is through single molecules or molecular
aggregates and has therefore to be treated quantum mechanically.  This
quantum nature makes it of course more complicated to determine, for
example, the current-voltage characteristics than in classical theories.
But at the same time a quantum treatment offers certain advantages and
especially the possibility of constructive or destructive interference
effects.

In the current letter we focus on the electron transport through a
molecular wire which is coupled to two leads acting as electron source and
drain. 
Many theories of quantum transport utilize the non-equilibrium Green's
function approach \cite{meir92} which is formally exact within the
lead-wire coupling. But because of the dependence of Green's functions on
two time arguments it is rather difficult to determine the current without
any further approximations, as for example, the wide-band limit.  Another
possible route is to treat the lead-wire coupling perturbatively and derive
quantum master equations for the electron dynamics in the wire and the
current through the wire \cite{kohl04a,li05,ovch05a,wela05a}.  In addition
to the wire-lead coupling a laser field can be coupled to the wire and/or
the leads. This would possibly allow for an ultrafast opto-electronic
device. First experimental \cite{gers00} and theoretical \cite{kohl04a,wela05a}
investigations in this direction have been performed. Most of the
theoretical studies are based on a master equation approach since in this
formalism allows easily to include time-dependent laser fields.  For
time-periodic fields the Floquet theory can be employed \cite{kohl04a}.
Recently the present authors derived a formalism which treats the
laser-wire interaction exactly for an arbitrary time-dependence of the
field \cite{wela05a}. This approach has been developed in analogy to the
treatments in dissipative quantum dynamics, but instead of a bosonic heat
bath in dissipation theories the leads are treated as fermionic reservoirs.
In addition, the system-bath coupling operator allows for particle exchange
in the case of molecular wires while in dissipative theories only energy
exchange is allowed. The present approach makes use of the spectral
decomposition of the spectral density $J(\omega)$ first introduced by Meier
and Tannor \cite{meie99}. These authors applied a time-nonlocal formalism
based on the Nakajima-Zwanzig equation. Later time-local theories based on
the same trick have been developed \cite{yan05,klei04a}. For molecular
wires these two versions of quantum master equations have been compared in
Ref.~\cite{wela05b}.

Grossmann et al.\cite{gros91,gros92} showed in their pioneering work that
periodically driven quantum systems can exhibit the unusual effect of
quenching the tunneling dynamics at specific values of field amplitude.
This effect was termed coherent destruction of tunneling (CDT). The
validity range of this localization of particles by external periodic
fields has later been studied in more detail for situations including a
bias\cite{stoc99}. For cw-laser fields CDT has already been described for
model molecular wires using the Floquet theory \cite{lehm03a}.  The purpose
of this work is to study the CDT for short Gaussian laser pulses
which have a high carrier frequency.  Earlier studies \cite{wela05a}
suggest that the CDT phenomenon survives also for shorter pulse length not
just infinite length as in the original CDT studies.  In those studies an
average current was investigated. For cw-laser fields the CDT appears to be
perfect if one measures at intervals separated by the period of the driving
field but there are still oscillations within one period.  Another way of
looking at the effect is by averaging over one period of the periodic driving
in the cw case. As will be discussed later, for shorter laser pulses one
needs to average over several of the high-frequency oscillations of the
carrier frequency.

\section{Model}

\begin{figure}
\onefigure[width=6cm]{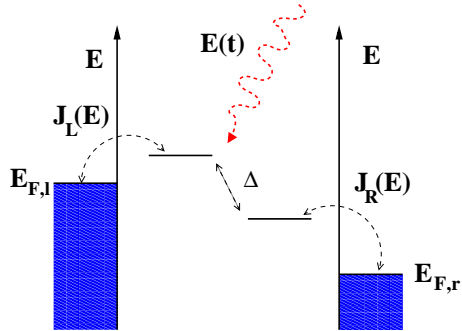}
\caption{Sketch of the two-site system defined in the text.}
\label{f.0}
\end{figure}

As in many quantum dissipation theories, the physical system is separated
into the relevant system $H_S(t)$, mimicking the wire, and reservoirs $H_R$
modeling the leads
\begin{equation} \label{equ:Ham_total}
H(t)=H_S(t)+H_R+H_{SR}
\end{equation}
with wire-lead coupling $H_{SR}$.  The
wire consists of $N$ sites coupled to each other by a hopping element
$\Delta$ (see Fig.~\ref{f.0}).  Denoting the creation (annihilation)
operator by $c^\dagger_n$ ($c_n$) the tight-binding description of the
electrons in the molecular wire reads
\begin{equation} \label{equ:Ham_wire}
H_S(t)=\sum_n E_n c_n^\dagger c_n - \Delta
\sum_{n} ( c_n^\dagger c_{n+1} +c_{n+1}^\dagger c_{n})
+ U \sum_{n}   c_n^\dagger c_{n}  c_{n+1}^\dagger c_{n+1}
- \mu{}E(t)~. 
\end{equation}
The first term describes the on-site energies, the second term the
nearest-neighbor hopping and the third term electron correlation within the
wire in a version for spinless particles. The fourth term gives the
coupling between the wire and the laser field $E(t)$ whereas for the dipole
operator we assume \cite{kohl04a}
\begin{eqnarray}
  \label{eq:1}
  \mu{}= e\sum_n x_n  = e \sum_n \frac{2n - N -1}{2}  c_n^\dagger c_n~.
\end{eqnarray}

The environment of the wire consists of two electronic leads that are
modeled by two independent reservoirs of uncorrelated electrons in thermal
equilibrium. For each lead, the Hamiltonian $H_R$ is given by
$ \label{equ:Ham_lead}
H_R=\sum_q \omega_q c_q^\dagger c_q $
with $c_q^\dagger$ and $c_q$ creating and annihilating an electron in the
corresponding reservoir mode $\vert q \rangle$ with  energy $\omega_q$. 
$\hbar$ is set to unity throughout the paper. Due
to the assumed thermal equilibrium of the electronic leads, the occupation
expectation values of the reservoir modes are determined by
$ \label{equ:equi}
\langle c_q^\dagger c_{q'} \rangle = n_F(\omega_q-E_F) \delta_{qq'}$ 
where $n_F$ is the Fermi function and $E_F$ the Fermi energy.  In further
derivations we will only refer to the left lead but the formalism has to be
applied to the right lead coupled to the last site $N$ of the wire as well.
The coupling of the left electronic lead with the first site of the wire is
given by
\begin{equation} \label{equ:Ham_coup}
H_{SR}= \sum_{x=1}^2 K_x  \Phi_x =  \sum_q (V_q c_1^\dagger c_q + V_q^* c_q^\dagger c_1)
\end{equation}
with $\Phi_1=\sum_q V_q c_q$, $\Phi_2=\sum_q V_q^* c_q^\dagger$,
$K_1=c_1^\dagger$, $K_2=c_1$, and a wire-lead coupling strength $V_q$ for
each reservoir mode.

As one is normally not interested in the dynamics within the leads but only
within the wire, a quantum master equation (QME) based on a second-order
perturbation theory in the wire-lead coupling has been developed
for the reduced density matrix of the wire $\rho_S(t)$ \cite{wela05a} 
\begin{equation}\label{equ:master2}
\dot{\rho}_S(t)=-i \mathcal L_S(t) \rho_S(t) - \sum_{xx'}^{1,2} [K_x,\, 
\Lambda_{xx'}(t)-\widehat \Lambda_{xx'} (t)]
\end{equation}
with $\mathcal L_S(t)$ being the Liouville operator applying $H_S$ and the auxiliary operators
\begin{equation}\label{equ:aux1}
\Lambda_{xx'}^{}(t) = \int_{t_0}^t \mathrm dt' C_{xx'}(t-t')  U_S(t,t')  K_{x'} \rho_S(t'),
\end{equation}
\begin{equation}\label{equ:aux2}
\widehat \Lambda_{xx'}(t) = \int_{t_0}^t \mathrm dt' C_{x'x}^*(t-t')  U_S(t,t') \rho_S(t') K_{x'}~.
\end{equation}
Here we employed the definitions $U_S(t,t')=T_+\exp\left(-i \int_{t'}^t
  \mathrm d\tau \mathcal L_S(\tau)\right)$ and the reservoir correlation
functions $C_{x x'}(t)=\mathrm{tr}_R \lbrace {\rm e}^{i H_R t} \Phi_x {\rm
  e}^{-i H_R t} \Phi_{x'} \rho_R \rbrace$ with the reservoir density matrix
$\rho_R$.  Using this form of the QME has the advantages over the standard
Redfield approach \cite{redf57} that the memory terms are included and that
the wire-laser coupling is still treated exactly within the dipole
approximation.

For the QME all the external properties of the fermionic reservoir are
described by a single quantity, namely the spectral density $
 \label{equ:spectralgeneral}
J_{R}(\omega)=\sum_q \pi \vert V_q \vert^2 \delta(\omega-\omega_q)$.
The sum becomes a smooth function for a dense
spectrum of the reservoir modes. Now we use the trick of a 
numerical decomposition of the spectral density \cite{meie99,wela05a}
\begin{equation} \label{equ:spectralnum}
J_{R}(\omega)=\sum_{k=1}^m \frac{p_k}{4 \Omega_k}  \frac{1}{(\omega -\Omega_k)^2+\Gamma_k^2},
\end{equation}
with real fitting parameters $p_k$, $\Omega_k$ and $\Gamma_k$. This
decomposition is not restricted to a any shape of the spectral density
and can therefore be applied to approximate complicated band structures. This
enables one to avoid the assumption of the wide-band limit and to take
influences of the band structure on the dissipative electron transfer
between the wire and the leads fully into account.

With the complex roots of the Fermi function and of function
(\ref{equ:spectralnum}), the application of the theorem of residues results
in
\begin{eqnarray} \label{bath12dec}
C_{12}(t)=\sum_{k=1}^m \frac{p_k}{4 \Omega_k \Gamma_k}
\left(n_F(-\Omega_k^- +E_F) e^{-i\Omega_k^- t} \right) 
-\frac{2i}{\beta} \sum_k^{m'} J_{R}(\nu_k^\ast) e^{-i \nu_k^\ast t}
=\sum_{k=1}^{m+m'} a_{12}^k e^{\gamma_{12}^k t}
\end{eqnarray}
\begin{eqnarray} \label{bath21dec}
C_{21}(t)= \sum_{k=1}^m \frac{p_k}{4 \Omega_k \Gamma_k}
\left(n_F(\Omega_k^+-E_F)  e^{i\Omega_k^+ t} \right) 
-\frac{2i}{\beta} \sum_k^{m'} J_{R}(\nu_k) e^{i \nu_k t}
=\sum_{k=1}^{m+m'} a_{21}^k e^{\gamma_{21}^k t}
\end{eqnarray}
with the abbreviations $\Omega_k^+=\Omega_k+i \Gamma_k$ and
$\Omega_k^-=\Omega_k-i \Gamma_k$ and the Matsubara frequencies
$\nu_k=i\frac{2\pi k + \pi}{\beta} +E_F$.  Rigorously, the sum over the
Matsubara frequencies would be infinite but it can be truncated at a finite
$m'$ depending on the temperature of the system $T$ and the spectral width
of $J_R(\omega)$.  The pure exponential dependence of the correlation
function on time allows one to derive a set of differential equations for
the auxiliary density operators
\begin{eqnarray}\label{equ:diffaux1}
\frac{\partial}{\partial t} \Lambda_{xx'}^k(t)&=& a_{xx'}^k  K_{x'} \rho_S(t) -i [H_S(t), \Lambda_{xx'}^k(t)] + \gamma_{xx'}^k  \Lambda_{xx'}^k(t),
\end{eqnarray}
\begin{eqnarray}\label{equ:diffaux2}
\frac{\partial}{\partial t}{\widehat\Lambda}_{xx'}^k(t)&=&\left(a_{x'x}^k\right)^\ast \rho_S(t) K_{x'}   -i [H_S(t), \widehat \Lambda_{xx'}^k(t)] + \left(\gamma_{x'x}^k \right)^\ast \widehat \Lambda_{xx'}^k(t)
\end{eqnarray}
with ${\Lambda}_{xx'}(t)=\sum_{k=1}^{m+m'} {\Lambda}_{xx'}^k(t)$ and
${\widehat\Lambda}_{xx'}(t)=\sum_{k=1}^{m+m'}
{\widehat\Lambda}_{xx'}^k(t)$.

As detailed in Ref.~\cite{wela05a} it is now easy to give an expression for
the current. Using the electron number operator of the left lead with
the summation performed over the reservoir degrees of freedom $N_l=\sum_q
c_q^{\dagger} c_q$ yields
\begin{equation}\label{equ:finalcurrent}
I_l(t)=e\frac{\mathrm d}{\mathrm dt} \mathrm{tr} \, \lbrace N_l \rho_S(t)
\rbrace =-ie \, \mathrm{tr} \left\{ [N_l,H(t)] \rho_S(t) \right\}
=2e \, \mathrm{Re} \left(\mathrm{tr}_S \left\{ c_1^\dagger \Lambda_{12}(t) -c_1^\dagger \widehat\Lambda_{12}(t) \right\} \right).
\end{equation}

\section{Numerical results}

In the following we set the value 0.1 eV for the tight-binding hopping
parameter $\Delta$. This parameter also constitutes the
energy scale for the results shown.  The system temperature is set to
$T=0.25 \Delta / k_B \approx 290$ K.  Though with the method described
above one can model complex energy dependences in the wire-lead coupling,
here we use for simplicity only one Lorentzian in the spectral density 
(\ref{equ:spectralnum}).  The maximum strength is chosen to be $0.1$
$\Delta$. This is achieved by using $p_1=0.644$ eV, $\Gamma_1=5.44$ eV, and $
\Omega_1=0.544$ eV.  With the chosen energy settings, a time unit in the
system corresponds to 0.66 fs. The resulting current unit can be extracted
from Eq.\ (\ref{equ:finalcurrent}) and is equal to a macroscopic value of
$1 [e]=2.43 *10^{-4}$~A. The carrier frequency is set to $1$ eV.

\begin{figure}
\twofigures[width=6cm]{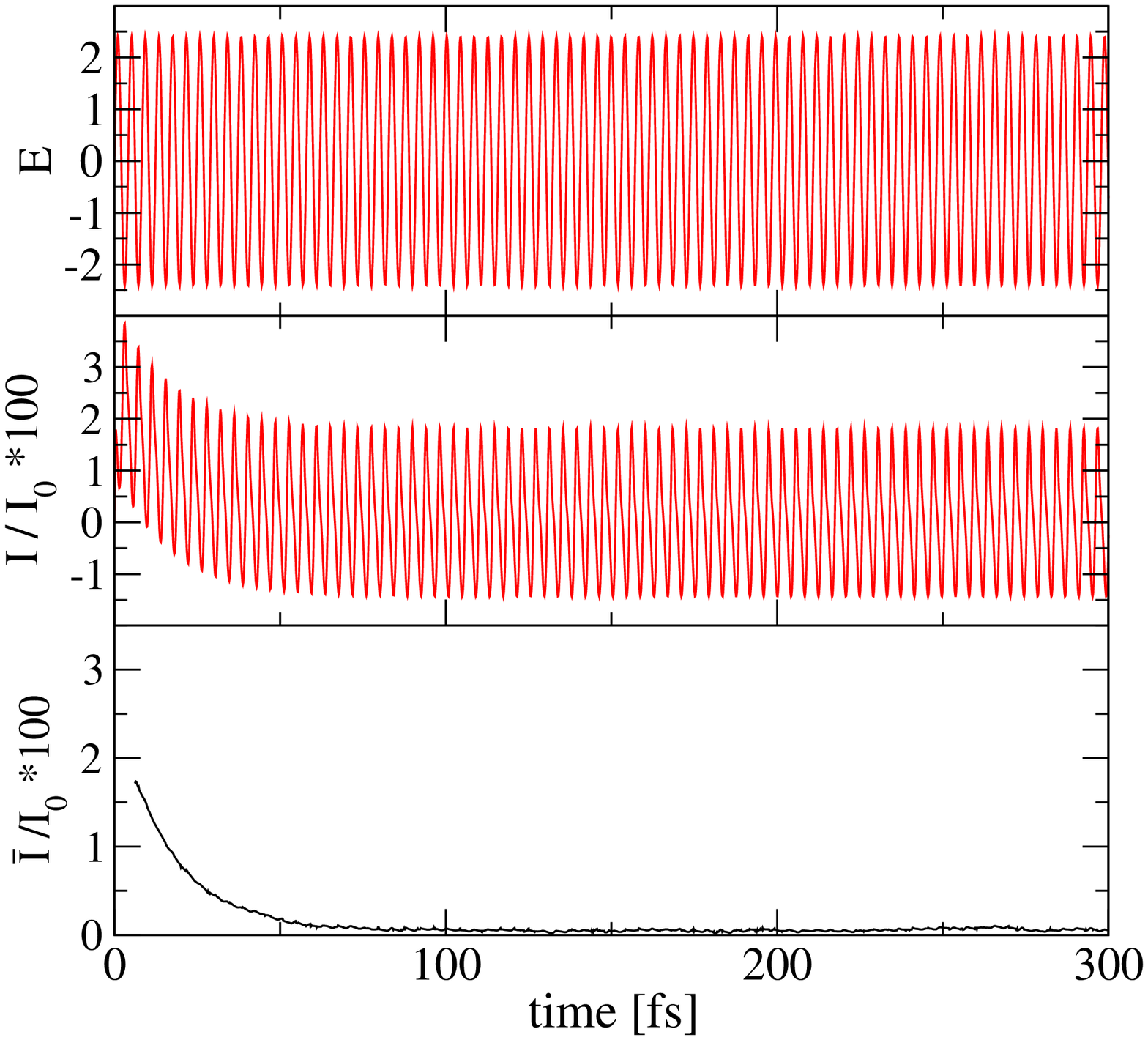}{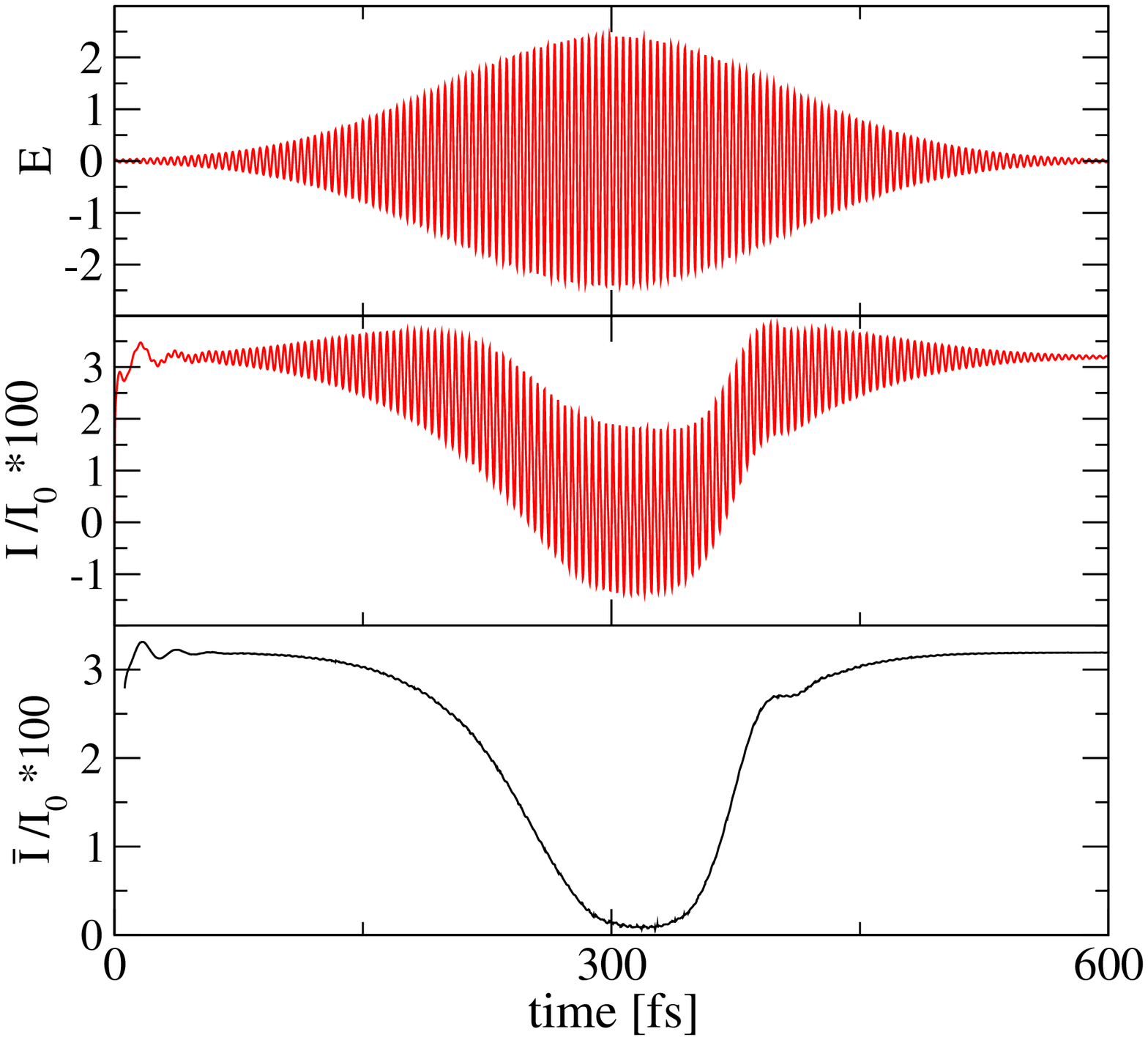}
\caption{The scenario of CDT for a monochromatic laser field. The top panel
shows the laser field, the middle one the current through the wire, and the
lowest panel the averaged current.}
\label{f.1}
\caption{Same as in \protect Fig.~\ref{f.1} but for a short laser pulse
  with length $\sigma$ = 50 fs.}
\label{f.2}
\end{figure}

In Fig.~\ref{f.1} the phenomenon of CDT is shown for a monochromatic laser
field of amplitude $A=2.405$ eV. This amplitude corresponds to a full CDT
\cite{lehm03a,wela05a}. The wire consists of two sites with on-site
energies $E_1=E_2=0.54$ eV which are initially empty. The Fermi energy of
the left lead is 1.0 eV and that of the right one is 0.0 eV.  The top panel
of Fig.~\ref{f.1} shows the laser field while the middle panel displays the
time-dependent current $I(t)$ through the wire. To be able to see the
effect more easily we also show an averaged time-dependent current
$\bar{I}$ determined by averaging $I(t)$ over three periods of the
highly-oscillating carrier field. Since the laser is instantaneously turned
on at $t=0$, one sees some transient behavior during the first 150 fs
before the CDT fully sets in.  From Fig.~\ref{f.2} one can deduce that the
transient behavior is mainly due to the slow onset of the CDT and not due
to filling the wire sites. In this figure no laser pulse is applied to the
wire for the first 150 fs and one observes a transient behavior during the
first 50 fs due to filling the wire sites. Then the laser pulse with a
Gaussian shape $E(t)=A \, \mathrm{exp} \left(-(t-T)^2/(2 \sigma^2) \right)
\mathrm{sin} \left(\omega_d t \right)$ with $\sigma=50$ fs and a maximum
amplitude as for the monochromatic laser before sets in.  For the short
laser pulse a complete suppression of the average current is not reached at
any time but a dip in the current is clearly visible. The shape of this
indentation does not show a Gaussian form nor is it symmetric. The current
decays smoothly towards a minimum value while at the end of the laser pulse
the current goes back to its equilibrium value showing some oscillations.

\begin{figure}
\twofigures[width=6cm]{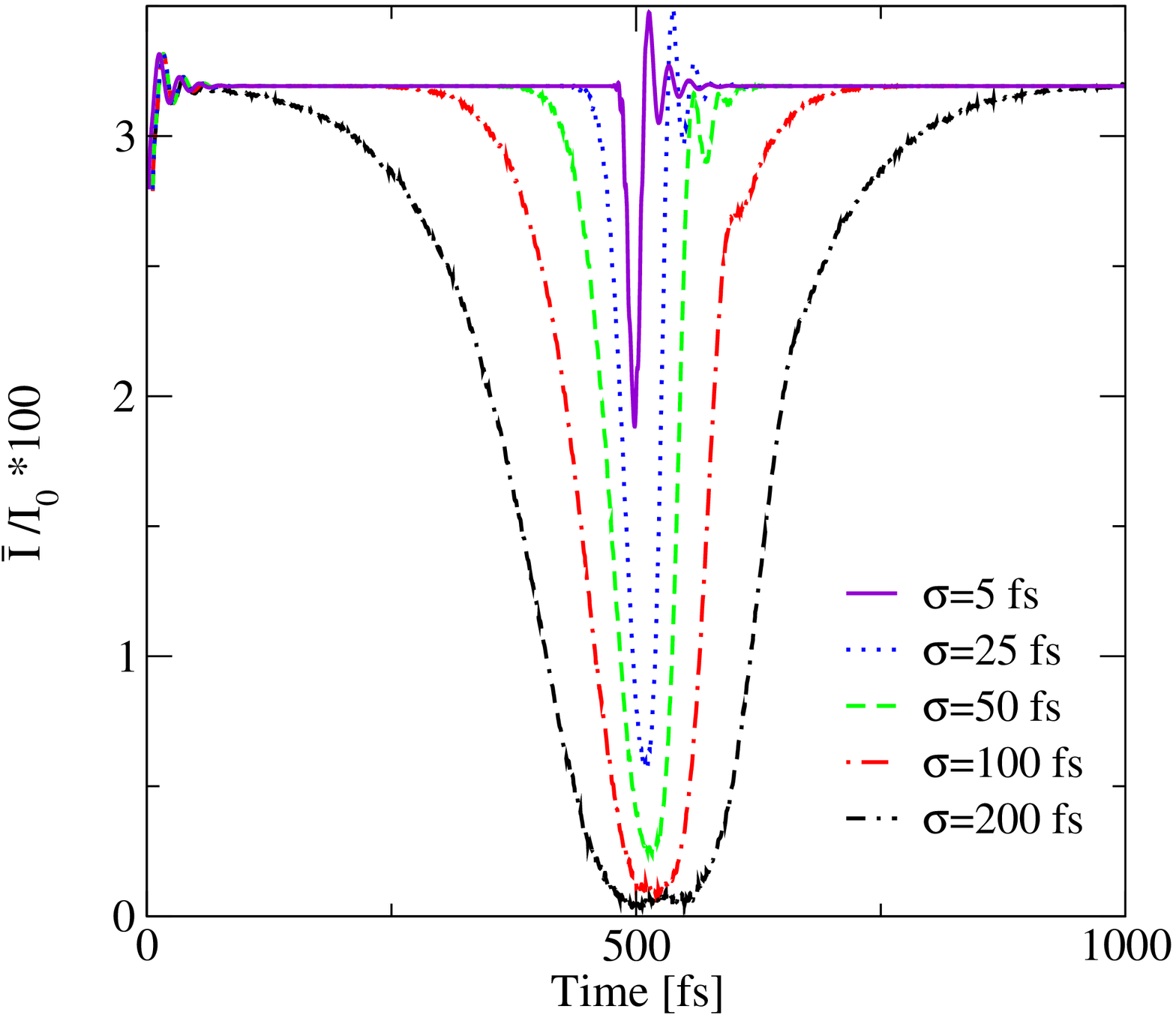}{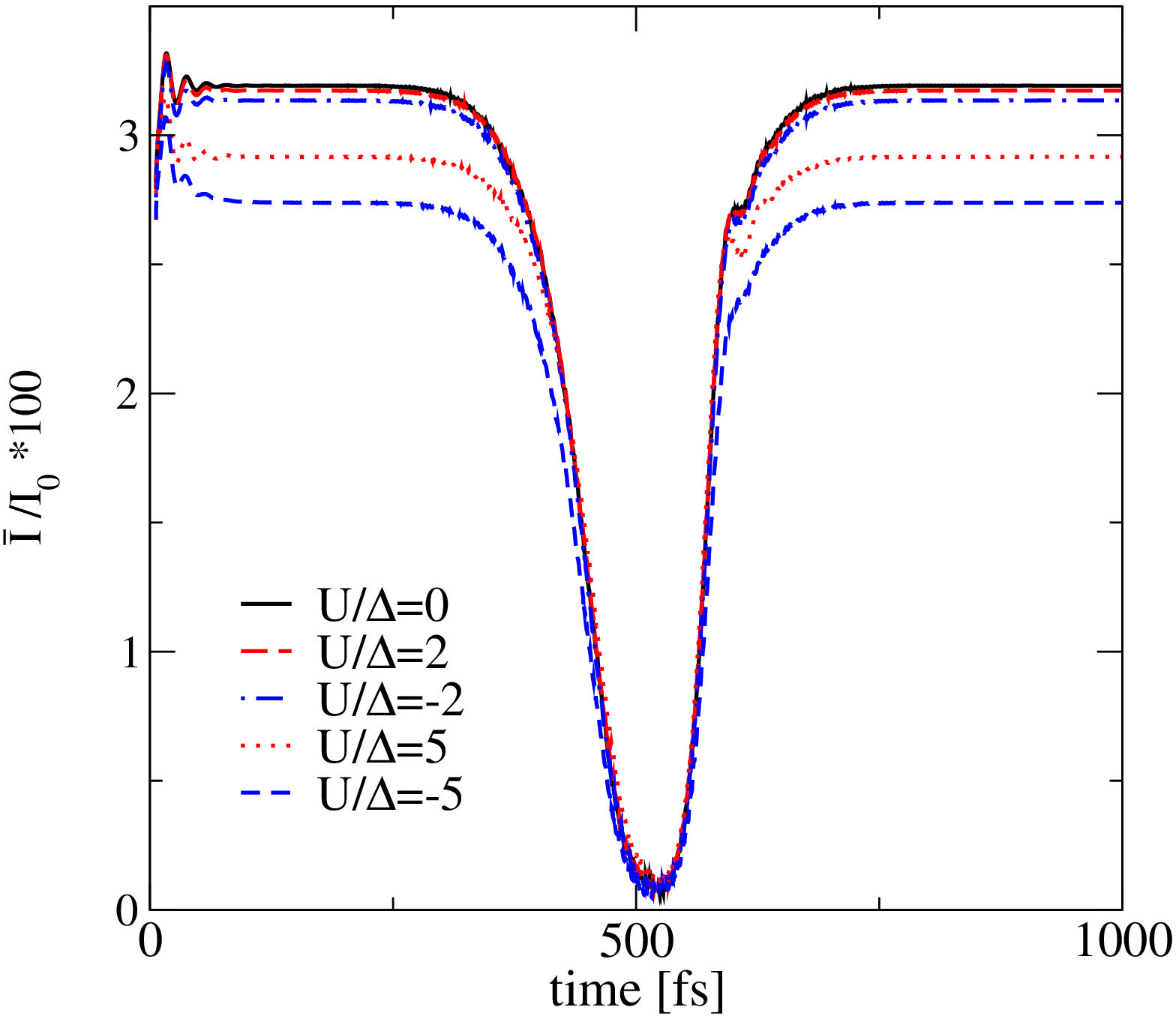}
\caption{The average current through a molecular wire for different pulse
  lengths. Other parameters are the same as in Fig.~3.}
\label{f.3}
\caption{The effect of electron
  correlation for a pulse with length $\sigma$ = 50 fs. Displayed are cases
  with no correlation, electron attraction and electron repulsion.}
\label{f.5}
\end{figure}
To investigate the dependence of the average current on the pulse length,
the average current is displayed in Fig.~\ref{f.3} for five different pulse
lengths. It is clearly visible that the minimum average current approaches
zero for longer pulse lengths. For an infinite pulse length the CDT
condition would be perfectly fulfilled and the average current would vanish
as shown in Fig.~\ref{f.1}. It is interesting that already for $\sigma=5$
fs one can see a clear dip in the average current although only very few
oscillations of the carrier field are involved in this scenario which reach
an amplitude close to the maximum for that pulse length.

\begin{figure}
\onefigure[width=6cm]{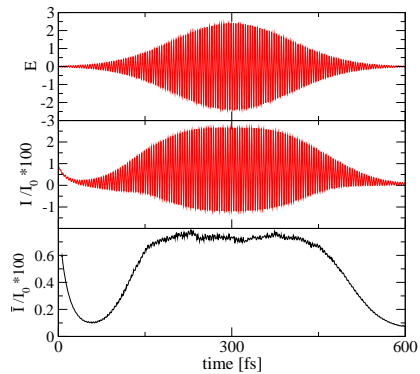}
\caption{After a short transient behavior no net current would flow through
  the wire for $E_{F,l}=0.82$ eV, $E_{F,r}=-0.82$ eV, $E_1=-0.82$ eV and
  $E_2=1.36$ eV.  The laser pulse which is the same as in Fig.~\ref{f.2}
  shifts the levels of the wire in such a manner that a temporary current is
  possible.}
\label{f.6}
\end{figure}

All the calculations above were performed for $U/\Delta=2$. To investigate
the effect of electron correlation on the CDT we also did the calculations
using a Gaussian laser pulse with $\sigma=50$ fs for different $U$. The
results in Fig.~\ref{f.5} show already an interesting transient behavior of
the current while the laser pulse is still off. Without correlation the
equilibrium current has the largest value. Turning on correlation the
equilibrium value of the average current decreases almost to the same value
for electron attraction and repulsion of $U/\Delta = \pm{}2$ .  For the larger
value of $U/\Delta=\pm{}5$ the equilibrium value of the current is reduced much
more for electron attraction than for repulsion. Interestingly enough, the
minimum value of the current is nearly the same in all cases regardless of
the strength and sign of correlation.  So the degree of CDT depends on the
length of the laser pulse but not on electron correlation.

Finally we want to show that effects of laser pulses are also visible if
the conditions for CDT are not fulfilled. Then in general a much weaker
suppression of the current is visible. But if one uses a configuration in
which, due to energetic reasons, a current is not possible but an external
bias is applied, then there are cases in which the laser can initiate a
current through the wire. This phenomenon can be seen in Fig.~\ref{f.6}.
The parameters are set so that the high energy level of the second wire
site blocks the current.  This creation of a current by a laser pulse does
not seem to follow a pattern as the CDT.  Its physical basis is quite
different: The laser pulse brings the levels of the wire to a position
which energetically allows for population transfer between the sites and
between leads and outermost sites. So this effect can simply be explained
by energetic reasoning and no interference effects are needed for an
interpretation.


\section{Summary and outlook}
The main point of this work is to show that the phenomenon of CDT exists in
molecular wires also for short laser pulses and not only for monochromatic
laser fields. Nevertheless the amplitude condition as for cw-laser fields
plays an important role. If these conditions are not fulfilled only a very
weak current suppression is achieved by the laser pulse. If the CDT
condition is fulfilled the current can be suppressed almost completely
already for short laser pulses. For longer pulses the current suppression
gets more effective. Electron correlation was treated in a version for
spinless fermions. Large correlation can have a rather large effect on the
equilibrium current, larger for electron attraction than repulsion. Finally
we showed that for parameter regions in which no equilibrium current
occurs, a laser pulse can lead to a weak current caused by temporary
shifting the energy levels of the wire into a position in which population
transfer is possible.

In conclusion we showed interesting possibilities how ultra-short laser
pulses can influence the current through molecular wires. Though the
present calculations were performed for a simple model system, the basic
physics should survive in more complex systems.  Here we have restricted
ourselves to a wire of only two sites although similar results can be
obtained for longer wires.  Using the effect of CDT with short laser pulses
allows for the construction of fast opto-electronic switches.  This opens,
of course, a huge variety of applications for similar scenarios and
possible technical devices even though a number of experimental hurdles
have to be passed before one can think of direct technical applications.



\begin{thebibliography}{10}

\bibitem{nitz03a}
\Name{Nitzan~A. \and Ratner~M.~A.}
\REVIEW {Science}{300}{2003}{1384}.

\bibitem{ghos04}
\Name{Ghosh~A.~W., Damle~P.~S., Datta~S., \and Nitzan~A.}
\REVIEW {MRS Bulletin}{6}{2004}{391}.

\bibitem{meir92}
\Name{Meir~Y. \and Wingreen~N.~S.}
\REVIEW {Phys. Rev. Lett.}{68}{1992}{2512}.

\bibitem{kohl04a}
\Name{Kohler~S., Lehmann~J., \and H\"anggi~P.}
\REVIEW {Phys. Rep.}{406}{2005}{379}.

\bibitem{li05}
\Name{Li~X.-Q., Luo~J.-Y., Yang~Y.-G., Cui~P., \and Yan~Y.~J.}
\REVIEW {Phys. Rev. B}{71}{2005}{205304}.

\bibitem{ovch05a}
\Name{Ovchinnikov~I.~V. \and Neuhauser~D.}
\REVIEW {J. Chem. Phys.}{122}{2005}{024707}.

\bibitem{wela05a}
\Name{Welack~S., Schreiber~M., \and Kleinekath\"ofer~U.}
\REVIEW {J. Chem. Phys.}{124}{2006}{044712}.

\bibitem{gers00}
\Name{Gerstner~V., Knoll~A., Pfeiffer~W., Thon~A., \and Gerber~G.}
\REVIEW {J. Appl. Phys.}{88}{2000}{4851}.

\bibitem{meie99}
\Name{Meier~C. \and Tannor~D.~J.}
\REVIEW {J. Chem. Phys.}{111}{1999}{3365}.

\bibitem{yan05}
\Name{Yan~Y.~J. \and Xu~R.-X.}
\REVIEW {Ann. Rev. Phys. Chem.}{56}{2005}{187}.

\bibitem{klei04a}
\Name{Kleinekath\"ofer~U.}
\REVIEW {J. Chem. Phys.}{121}{2004}{2505}.

\bibitem{wela05b}
\Name{Welack~S., Kleinekath\"ofer~U., \and Schreiber~M.}
\newblock Laser-driven molecular wires studied by a non-markovian density
  matrix approach
\Year{2006}.
\newblock \mbox{J.} Lumin.\ (accepted).

\bibitem{gros91}
\Name{Grossmann~F., Dittrich~T., Jung~P., \and H\"anggi~P.}
\REVIEW {Phys. Rev. Lett.}{67}{1991}{516}.

\bibitem{gros92}
\Name{Grossmann~F. \and H\"anggi~P.}
\REVIEW {Europhys. Lett.}{18}{1992}{571}.

\bibitem{stoc99}
\Name{Stockburger~J.~T.}
\REVIEW {Phys. Rev. E}{59}{1999}{R4709}.

\bibitem{lehm03a}
\Name{Lehmann~J., Camalet~S., Kohler~S., \and H\"anggi~P.}
\REVIEW {Chem. Phys. Lett.}{368}{2003}{282}.

\bibitem{redf57}
\Name{Redfield~A.~G.}
\REVIEW {IBM J. Res. Dev.}{1}{1957}{19}.

\end{thebibliography}

\end{document}